\documentclass[11pt]{article}
\usepackage[extrawidemargins]{layout}
\usepackage{times-lite}
\usepackage[tbtags]{amsmath}
\usepackage{textcomp}
\usepackage{amssymb}
\usepackage{amsthm}
\usepackage{bm}
\usepackage{pl}
\usepackage{stmaryrd}
\usepackage{url}
\usepackage{supertabular}
\usepackage{float}
\usepackage{cite}
\usepackage{hyperref}
\hypersetup{
   pdftitle={Nexus Authorization Logic (NAL): Logical Results},
   pdfauthor={Michael R. Clarkson and Andrew K. Hirsch}
 }
\usepackage{defs}

\newcommand{\nalTrue}{\mathsf{true}}
\newcommand{\nalFalse}{\mathsf{false}}
\newcommand{\nalOr}{\vee}
\newcommand{\nalAnd}{\wedge}
\newcommand{\nalImplies}{\Rightarrow}
\newcommand{\nalNot}{\neg}
\newcommand{\nalForall}[2]{\forallqer{#1}{#2}}
\newcommand{\nalExists}[2]{\existsqer{#1}{#2}}
\newcommand{\NALSAYS}{\mathsf{says}}
\newcommand{\nalSays}[2]{{#1}\mathrel{\NALSAYS}{#2}}
\newcommand{\nalSubprin}[2]{{#1}.{#2}}
\newcommand{\nalGroup}[2]{\{{#1}\mathrel{:}{#2}\}}
\newcommand{\NALSPEAKSFOR}{\Rrightarrow}
\newcommand{\nalSpeaksfor}[2]{{#1}\NALSPEAKSFOR{#2}}
\newcommand{\nalSpeaksforRest}[4]{{#1}\Rrightarrow{#2}\mathrel{\mathsf{on}}(#3 : #4)}
\newcommand{\provesJ}[2]{{#1}\vdash{#2}}
\newcommand{\subst}[2]{[#1/#2]}
\newcommand{\FV}{\mathit{FV}}

\newcommand{\oldnal}{{NAL$_0$}}
\newcommand{\newnal}{{NAL$_1$}}

\begin{document}

\title{Nexus Authorization Logic (NAL): \\ Logical Results}
\author{Andrew K. Hirsch \quad Michael R. Clarkson  \\
  Department of Computer Science \\
  George Washington University \\
  $\{$akhirsch, clarkson$\}$@gwu.edu
}
\date{November 15, 2012}
\maketitle

\begin{abstract}
Nexus Authorization Logic (NAL) [Schneider et al. 2011]
is a logic for reasoning
about authorization in distributed systems.  A revised version
of NAL is given here, including revised syntax, a revised
proof theory using localized hypotheses, and a new
Kripke semantics.  The proof theory is proved sound
with respect to the semantics, and that proof is formalized in Coq.
\end{abstract}

\section{Introduction}

Authorization logics are epistemic logics used to reason
about whether principles are permitted to take actions in
a distributed computer system.  
Nexus Authorization Logic (NAL), invented by 
Schneider et al.~\cite{SchneiderWS11},
is notable for enabling rich reasoning about 
axiomatic, synthetic, and analytic bases for authorization
of actions.  NAL
extends a well-known authorization logic, 
cut-down dependency core calculus (CDD)~\cite{Abadi07}.
Among other features, NAL upgrades CDD from
having only propositional variables to having 
functions and predicates on system state. 

The NAL rationale~\cite{SchneiderWS11} gives
a natural-deduction proof system for the logic and
sketches the intuition for a semantics based on the idea
of a \emph{worldview}, which is the set of statements
that a principle believes, or would be prepared to support.
However, neither a formal semantics nor a proof of soundness
is given in the rationale.

Here, we initiate the formal study of the metatheory of NAL 
by developing a formal semantics and a proof of soundness.
Along the way, we streamline NAL in various ways, particularly
in the syntax (by eliminating second-order quantification) 
and in the proof system (by localizing hypothetical judgments).
We also fix a bug in the original proof system, which allowed
derivation of a formula that arguably should be considered
invalid.

Since our formalization of NAL differs from
that of the NAL rationale,
it will be convenient to have names to distinguish
these two formal systems.  Henceforth, we write
``{\oldnal}'' to refer to the original formalization of the
NAL rationale~\cite{SchneiderWS11}, and ``{\newnal}'' to refer to the new
formalization in this paper.

Our proof of soundness, including the syntax, proof
system, and semantics of {\newnal}, is formalized in the Coq proof
assistant.\footnote{\url{http://coq.inria.fr}} The 
formalization contains about 3,000 lines of code.

This short paper describes our formal syntax,
proof system, and semantics for NAL.  Familiarity with 
epistemic logics, constructive logics, and their 
Kripke semantics is assumed.  Readers who seek
background in these areas can consult standard
references~\cite{FaginHMV95,TroelstravD88}.

\section{Syntax}

{\newnal} is a constructive, first order, multimodal logic.  
It has two syntactic classes, terms $\tau$ and
formulas $\phi$.  Metavariable $x$ ranges over 
first-order variables, $f$ over first-order functions, 
and $r$ over first-order relations.  Logical formulas $\phi$ are
described by the following grammar:
\begin{align*}
\phi ::= 
\quad&\nalTrue \\
\bnf &\nalFalse \\
\bnf &r(\tau, \ldots, \tau) &\text{first-order relation} \\
\bnf &\tau_1 = \tau_2 &\text{term equality}\\
\bnf &\phi_1 \nalAnd \phi_2 &\text{conjunction} \\
\bnf &\phi_1 \nalOr \phi_2 &\text{disjunction} \\
\bnf &\phi_1 \nalImplies \phi_2 &\text{implication} \\
\bnf &\nalNot \phi &\text{negation} \\
\bnf &\nalForall{x}{\phi} &\text{first-order universal quantification} \\
\bnf &\nalExists{x}{\phi} &\text{first-order existential quantification}\\
\bnf &\nalSays{\tau}{\phi} &\text{affirmation} \\
\bnf &\nalSpeaksfor{\tau_1}{\tau_2}&\text{delegation} \\
\bnf &\nalSpeaksforRest{\tau_1}{\tau_2}{x}{\phi} 
  &\text{restricted delegation}
\end{align*}
Unlike {\oldnal}, formulas of {\newnal} do not permit monadic second-order
universal quantification.  In {\oldnal}, that quantifier was used only
to define certain connectives, particularly delegation, 
as syntactic sugar.  {\newnal} 
instead adds delegation as a primitive connective to the logic.
This simplifies the logic 
from second-order down to first-order,
at the small cost of adding a few extra axioms to the proof system
to handle the delegation primitive.

Logical terms are described by the following grammar:

\begin{align*}
\tau ::= 
\quad&x &\text{first-order variable}\\
\bnf &f(\tau, \ldots, \tau) &\text{first-order function}\\
\bnf &\nalSubprin{\tau_1}{\tau_2} &\text{subprincipal}\\
\bnf &\nalGroup{x}{\phi} &\text{group principal}
\end{align*}

There are some small, unimportant syntactic 
differences between {\oldnal} and {\newnal}.  The biggest of these
is the notation for delegation:  {\oldnal} uses $\rightarrow$, whereas 
{\newnal} uses $\NALSPEAKSFOR$ to avoid any potential confusion
with implication.

\section{Proof System}

The {\newnal} proof system is a natural-deduction proof system, like
the {\oldnal} proof system.  But unlike the {\oldnal} proof system,
which uses hypothetical judgments for proving implication introduction,
the {\newnal} proof system uses localized hypotheses.

In {\newnal}, the derivability judgment 
is written $$\provesJ{\Gamma}{\phi}$$
where $\Gamma$ is a set of formulas.  If $\provesJ{\Gamma}{\phi}$,
then $\phi$ is derivable from $\Gamma$ according to the rules
of the proof system.
Rules for formulas are given in figure~\ref{fig:rulesformulas}.
Rules for terms are given in figure~\ref{fig:rulesterms}.
In those figures, $\phi\subst{\tau}{x}$ denotes
capture-avoiding substitution of $\tau$ for $x$ in $\phi$.

Most of the proof system is routine.  The rules for $\NALSAYS$
use notation $\nalSays{p}{\Gamma}$, which intuitively means
that $p$ says all the formulas in set $\Gamma$.  Formally,
$\nalSays{p}{\Gamma}$ is the set
$\setdef{\nalSays{p}{\phi}}{\phi \in \Gamma}$.
The $\NALSAYS$ rules necessarily differ from the corresponding
rules found in {\oldnal} because of the use of localized 
hypotheses $\Gamma$.  Nonetheless, the {\newnal} rules
are essentially standard---for example, two of the
three rules correspond
to standard natural deduction rules for a necessity
modality~\cite{HughesC96}, and the third rule is symmetric
to the second.  

There is one important, deliberate change in the {\newnal}
proof system that makes its theory differ from the {\oldnal} system,
which we now discuss.  There are two standard ways of importing
beliefs into a principal's worldview.  The first is a rule known as 
Necessitation:
``if $\provesJ{}{p}$ 
then $\provesJ{}{\nalSays{p}{\phi}}$.''
The second is an axiom known as Unit:
$\provesJ{}{p \nalImplies (\nalSays{p}{\phi})}$.
Though superficially similar, Necessitation
and Unit lead to different theories.


\begin{example}
Machines $M_1$ and $M_2$ execute processes $P_1$ and $P_2$, respectively.   
$M_1$ has a register $R$.  Let $Z$ be a proposition representing 
``register $R$
is currently set to zero.''  According to Unit, $\provesJ{}{Z \nalImplies 
(\nalSays{P_1}{Z})}$ and $\provesJ{}{Z \nalImplies (\nalSays{P_2}{Z})}$.  The
former means that a process on a machine knows the current contents of
a register on that machine; the latter means that a process on a different
machine must also know the current contents of the register. 
But according to Necessitation, if $\provesJ{}{Z}$ then 
$\provesJ{}{\nalSays{P_1}{Z}}$ and $\provesJ{}{\nalSays{P_2}{Z}}$.  Only if
$R$ is always zero must the two processes say so.  
\end{example}

Unit, therefore, is better used when propositions (or relations or functions)
represent global state upon which all principals are guaranteed to agree.
Necessitation is better used when propositions represent 
local state that could be unknown to some principals.

NAL was designed to reason about state in distributed systems,
where principals (such as machines) may have local state,
and where global state does not necessarily exist---the reading at
a clock, for example, is not agreed upon by all principals 
in NAL. So Unit would be an overly strong restriction on NAL principals; 
Necessitation is the appropriate choice.  Fortunately,
{\oldnal} does include Necessitation as an inference rule and does not
include Unit as an axiom.

Unfortunately, {\oldnal}~\cite{SchneiderWS11} permits
Unit to be derived as a 
theorem\footnote{From $\mathcal{F}$ infer $\nalSays{A}{\mathcal{F}}$
by \textsc{says-i}.  Then infer 
$\mathcal{F} \nalImplies \nalSays{A}{\mathcal{F}}$ by \textsc{imp-i}.}
because of an interaction
between Necessitation and the {\oldnal} introduction rule for implication.
{\newnal} fixes this bug and does not permit derivation of Unit.

\begin{figure}[p]
\begin{equation*}
\renewcommand{\arraystretch}{3}
\begin{array}{cc}
\infer{\provesJ{\Gamma,\phi}{\phi}}{}
&
\infer{\provesJ{\Gamma,\psi}{\phi}}{\provesJ{\Gamma}{\phi}}
\\
\infer{\provesJ{\Gamma}{\nalTrue}}{}
&
\infer{\provesJ{\Gamma}{\phi}}{\provesJ{\Gamma}{\nalFalse}}
\\
\infer{\provesJ{\Gamma}{\phi \nalAnd \psi}}{\provesJ{\Gamma}{\phi} & \provesJ{\Gamma}{\psi}}
&
\infer{\provesJ{\Gamma}{\phi}}{\provesJ{\Gamma}{\phi \nalAnd \psi}}
\quad
\infer{\provesJ{\Gamma}{\psi}}{\provesJ{\Gamma}{\phi \nalAnd \psi}}
\\
\infer{\provesJ{\Gamma}{\phi_1 \nalOr \phi_2}}{\provesJ{\Gamma}{\phi_1}}
\quad
\infer{\provesJ{\Gamma}{\phi_1 \nalOr \phi_2}}{\provesJ{\Gamma}{\phi_2}}
&
\infer{\provesJ{\Gamma}{\psi}}{\provesJ{\Gamma}{\phi_1 \nalOr \phi_2} & \provesJ{\Gamma,\phi_1}{\psi} & \provesJ{\Gamma,\phi_2}{\psi}}
\\
\infer{\provesJ{\Gamma}{\phi \nalImplies \psi}}{\provesJ{\Gamma,\phi}
{\psi}}
& 
\infer{\provesJ{\Gamma}{\psi}}{\provesJ{\Gamma}{\phi} & \provesJ{\Gamma}{\phi \nalImplies \psi}}
\\
\infer{\provesJ{\Gamma}{\nalNot\phi}}{\provesJ{\Gamma,\phi}{\nalFalse}}
&
\infer{\provesJ{\Gamma}{\nalFalse}}{\provesJ{\Gamma}{\phi} & \provesJ{\Gamma}{\nalNot\phi}}
\\
\infer{\provesJ{\Gamma}{\nalForall{x}{\phi}}}{\provesJ{\Gamma}{\phi} & x \not\in \FV(\Gamma)}
&
\infer{\provesJ{\Gamma}{\phi\subst{\tau}{x}}}{\provesJ{\Gamma}{\nalForall{x}{\phi}}}
\\
\infer{\provesJ{\Gamma}{\nalExists{x}{\phi}}}{\provesJ{\Gamma}{\phi\subst{\tau}{x}}}
&
\infer{\provesJ{\Gamma}{\psi}}{\provesJ{\Gamma}{\nalExists{x}{\phi}} & \provesJ{\Gamma,\phi}{\psi} & x \not\in \FV(\Gamma,\psi)}
\\
\multicolumn{2}{c}{
\infer{\provesJ{\nalSays{p}{\Gamma}}{\nalSays{p}{\phi}}}{\provesJ{\Gamma}{\phi}}
\quad
\infer{\provesJ{\nalSays{p}{\Gamma}}{\nalSays{p}{\phi}}}{\provesJ{\nalSays{p}{\Gamma}}{\phi}}
\quad
\infer{\provesJ{\nalSays{p}{\Gamma}}{\nalSays{p}{\phi}}}{\provesJ{\Gamma}{\nalSays{p}{\phi}}}
}
\end{array}
\end{equation*}
\caption{Derivability judgment for formulas\label{fig:rulesformulas}}
\end{figure}

\begin{figure}[p]
\begin{equation*}
\renewcommand{\arraystretch}{3}
\begin{array}{c}
\infer{\provesJ{\Gamma}{\tau = \tau}}{}
\quad
\infer{\provesJ{\Gamma}{\tau_2 = \tau_1}}{\provesJ{\Gamma}{\tau_1 = \tau_2}}
\quad
\infer{\provesJ{\Gamma}{\tau_1 = \tau_3}}{\provesJ{\Gamma}{\tau_1 = \tau_2} & \provesJ{\Gamma}{\tau_2 = \tau_3}}
\\
\infer{\provesJ{\Gamma}{f(\tau_1,\ldots,\tau_n) = f(\tau'_1,\ldots,\tau'_n)}}{\provesJ{\Gamma}{\tau_1 = \tau'_1} & \ldots & \provesJ{\Gamma}{\tau_n = \tau'_n}}
\\
\infer{\provesJ{\Gamma}{r(\tau'_1,\ldots,\tau'_n)}}{\provesJ{\Gamma}{r(\tau_1,\ldots,\tau_n)} & \provesJ{\Gamma}{\tau_1 = \tau'_1} & \ldots & \provesJ{\Gamma}{\tau_n = \tau'_n}}
\\
\infer{\provesJ{\Gamma}{\nalSpeaksfor{\tau_1}{\tau_2}}}{\provesJ{\Gamma}{\nalSays{\tau_2}{\nalSpeaksfor{\tau_1}{\tau_2}}}}
\quad
\infer{\provesJ{\Gamma}{\nalSpeaksforRest{\tau_1}{\tau_2}{x}{\phi}}}{\provesJ{\Gamma}{\nalSays{\tau_2}{\nalSpeaksforRest{\tau_1}{\tau_2}{x}{\phi}}}}
\\
\infer{\provesJ{\Gamma}{\nalSays{\tau_2}{\phi}}}{\provesJ{\Gamma}{\nalSpeaksfor{\tau_1}{\tau_2}} & \provesJ{\Gamma}{\nalSays{\tau_1}{\phi}}}
\quad
\infer{\provesJ{\Gamma}{\nalSays{\tau_2}{\phi\subst{\tau}{x}}}}{\provesJ{\Gamma}{\nalSpeaksforRest{\tau_1}{\tau_2}{x}{\phi}} & \provesJ{\Gamma}{\nalSays{\tau_1}{\phi\subst{\tau}{x}}}}
\\
\infer{\provesJ{\Gamma}{\nalSpeaksfor{\tau}{\tau}}}{}
\quad
\infer{\provesJ{\Gamma}{\nalSpeaksforRest{\tau}{\tau}{x}{\phi}}}{}
\\
\infer{\provesJ{\Gamma}{\nalSpeaksfor{\tau_1}{\tau_3}}}{\provesJ{\Gamma}{\nalSpeaksfor{\tau_1}{\tau_2}} & \provesJ{\Gamma}{\nalSpeaksfor{\tau_2}{\tau_3}}}
\quad
\infer{\provesJ{\Gamma}{\nalSpeaksforRest{\tau_1}{\tau_3}{x}{\phi}}}{\provesJ{\Gamma}{\nalSpeaksforRest{\tau_1}{\tau_2}{x}{\phi}} & \provesJ{\Gamma}{\nalSpeaksforRest{\tau_2}{\tau_3}{x}{\phi}}}
\\
\infer{\provesJ{\Gamma}{\nalSpeaksfor{\tau}{\nalGroup{x}{\phi}}}}{\provesJ{\Gamma}{\phi\subst{\tau}{x}}}
\quad
\infer{\provesJ{\Gamma}{\nalSpeaksfor{\nalGroup{x}{\phi}}{\tau}}}{\provesJ{\Gamma,\phi}{\nalSpeaksfor{x}{\tau}} & x \not\in\FV(\tau)}
\\
\infer{\provesJ{\Gamma}{\nalSpeaksfor{\tau_1}{\nalSubprin{\tau_1}{\tau_2}}}}{}
\quad
\infer{}{}
\end{array}
\end{equation*}
\caption{Derivability judgment for terms\label{fig:rulesterms}}
\end{figure}

\section{Semantics}

The semantics of {\newnal} is combination of three 
standard semantic models:  first-order models, constructive models,
and modal models.  This combination is probably not completely
novel (see, e.g., \cite{Wijesekera90,GenoveseGR12}), though we are not aware
of any authorization logic semantics 
that is identical to or that subsumes our semantics. 
Our presentation mostly follows the Kripke semantics of intuitionistic
predicate calculus given by Troelstra and van Dalen~\cite{TroelstravD88}.

Below, we give a moderately pedagogic description of the definition of a
semantic model for NAL, by building up progressively more complicated models.

\paragraph*{First-order models.} 
A \emph{first-order model with equality} is a tuple $(D, =, R, F)$.
The purpose of a first-order model is to interpret the first-order
fragment of the logic, specifically first-order quantification,
functions, and relations.  
$D$ is a set, the \emph{domain} of individuals.  These individuals
are what quantification in the logic ranges over.  
$R$ is a set $\setdef{r_i}{i \in I}$ of relations on $D$,
indexed by set $I$, with associated arity function
$m$, such that $r_i \subseteq D^{m(i)}$.
Likewise, $F$ is a set $\setdef{f_j}{j \in J}$ of functions on $D$, 
indexed by set $J$, with associated arity function
$n$, such that $f_j \in D^{n(j)} \rightarrow D$.
There is a distinguished equality relation $=$, which 
is an equivalence relation on $D$, such that 
equality is indistinguishable by relations and functions:
\begin{itemize}
\item if $\vec{d} = \vec{d'}$ and $\vec{d} \in r_i$, where 
  $|\vec{d}| = |\vec{d'}| = m(i)$, then $\vec{d'} \in r_i$, and
\item if $\vec{d} = \vec{d'}$, where 
  $|\vec{d}| = |\vec{d'}| = n(j)$, then $f_j(\vec{d}) = f_j(\vec{d'})$.
\end{itemize}

\paragraph*{Constructive models.}
A \emph{constructive model} is a tuple $(W, \leq, s)$.  The purpose
of a constructive model is to interpret the constructive fragment
of the logic, specifically implication and universal quantification,
(whose semantics differ from the classical semantics).
$W$ is a set, the \emph{possible worlds}. We denote
an individual world as $w$.  Intuitively, a world $w$ represents
the \emph{state of knowledge} of a constructive reasoner.
Relation $\leq$, called the \emph{constructive
accessibility relation}, is a partial order on $W$. If $w \leq w'$, then
the constructive reasoner's state of knowledge could grow from $w$ to
$w'$. Function $s$ is called the \emph{interpretation function}. It assigns
a first-order model $({D_w}, {=_w}, {R_w}, {F_w})$ to each world $w$.
(Let the individual elements of $R_w$ be notated as 
$\setdef{r_{i,w}}{i \in I}$, and likewise for $F_w$, as
$\setdef{f_{j,w}}{j \in J}$.)  Thus, $s$ enables
a potentially different first-order interpretation at each world.  
But to help ensure that the constructive reasoner's state of knowledge
only grows---hence never invalidates a previously admitted
construction---we require $s$ to be
monotonic w.r.t.\ $\leq$.  That is, if $w \leq w'$ then
\begin{itemize}
\item $D_w \subseteq D_{w'}$,
\item $d =_w d'$ implies $d =_{w'} d'$,
\item $r_{i,w} \subseteq r_{i,w'}$, and
\item for all $\vec{d}$ such that $|\vec{d}| = n(j)$, it holds that
$f_{j,w}(\vec{d}) =_w f_{j,w'}(\vec{d})$.
\end{itemize}

\paragraph*{Constructive modal models.} 
A \emph{constructive modal model} is a tuple
$(W, \leq, s, P, A)$.
The purpose of a constructive modal model is to interpret the 
modal fragment of the logic, specifically the 
$\NALSAYS$ connective and the delegation connectives.
The first part of a constructive modal model, $(W, \leq, s)$, must itself
be a constructive model as above.  The next part, $P$, is a 
set of \emph{principals}.  Note that we treat principals differently
than individuals:  although individuals can vary from
world to world in a model, 
the set of principals is assumed to be constant across the entire model.
This assumption is consistent with other constructive multimodal 
logics~\cite{Wijesekera90,Simpson94}, which have a fixed set of 
modalities (just $\Box$ and $\Diamond$).
However, it would be interesting in future work to explore
removing this assumption.

$A$ is a set $\setdef{A_p}{p \in P}$ of binary 
relations on $W$, called the \emph{principal accessibility relations}.
If $(w,w') \in A_p$, then in world $w$, principal $p$ considers 
world $w'$ possible. Like $\leq$ in a constructive model,
we require $s$ to be monotonic w.r.t.\ each $A_p$.
This requirement 
enforces a kind of constructivity on each principal $p$, such that 
if $p$ is in a world in which individual $d$ is constructed, then
$p$ cannot consider possible any world in which $d$ has not
been constructed.

Constructive modal models thus have two kinds of accessibility
relations, constructive $\leq$ and principal $A_p$.  These
relations cannot be completely orthogonal: for sake of soundness,
we need to impose four \emph{frame conditions} that relate
constructive accessibility and principal accessibility.
\begin{itemize}
\item \textbf{F1.} If $w \leq w'$ and $(w, v) \in A_p$, then
there exists a $v'$ such that $v \leq v'$ and $(w', v') \in A_p.$
\item \textbf{F2.} If $(w, v) \in A_p$ and $v \leq v'$, then
there exists a $w'$ such that $w \leq w'$ and $(w', v') \in A_p.$
\item \textbf{IT.} If $(w, v) \in A_p$ and $(v, u) \in A_p$,
then there exists a $w'$ such that $w \leq w'$ and $(w', u) \in A_p$.
\item \textbf{ID.} If $(w, u) \in A_p$, then there exists
a $w'$ and $v$ such that $w \leq w'$, and $(w', v) \in A_p$ as well as
$(v, u) \in A_p$.
\end{itemize}
The need for these frame conditions originates from the proof system
rules for $\NALSAYS$, especially the latter two rules.  It's well
known in modal logic that axioms and rules about modalities
correspond to frame conditions on accessibility relations 
(see, e.g., chapter 3 of~\cite{FaginHMV95}).  IT and ID
are intuitionistic generalizations of transitivity and density
of the $A_p$ relations.  In the presence of F1 and F2, IT and ID
are necessary and sufficient conditions for the soundness
of the $\NALSAYS$ rules---a result that follows from work
by Plotkin and Stirling~\cite{PlotkinS86}.  Furthermore,
F1 and F2 are arguably the right fundamental frame
conditions to impose in a constructive modal logic~\cite{Simpson94}.
In the case of NAL, we could actually remove F1 without suffering
any unsoundness or incompleteness.  (F1 is needed only
to show soundness of a $\Diamond$ modality, which does not
exist in NAL.)  However, the others---F2, IT, and ID---are
all necessary to impose in {\newnal}.

\paragraph*{NAL models.} 
A \emph{NAL model} is a tuple
$(W, \leq, s, P, A, \vee, \bot, \mathit{SUB})$.
The purpose of a NAL model is to interpret NAL formulas.  
Specifically, it adds machinery to interpret group and subprincipals.
The first part of a NAL model, $(W, \leq, s, P, A)$, must itself
be a constructive modal model as above.  

The next part of a NAL
model, $(\vee, \bot)$, is used to interpret group principals.  
Specifically, $(P, \vee)$ must be a join semilattice, with $\bot$ as
its bottom element.  (Thus, $\bot$ is a principal.  Its intended
use is as a principal who believes only tautologies.  We do not
require the existence of a top element in the lattice, because
there is no need for such an element in the semantics.)  Join operator
$\vee$ is used to take disjunctions of principals---intuitively,
$p \vee q$ is the principal who believes those statements
that either $p$ or $q$ believe, or statements that logically
follow from those.  Formally, we require that, for all principals $p$
and $q$, it holds that $A_{p \vee q} \subseteq A_p$.

The $\mathit{SUB}$ part of a NAL model is used to interpret
subprincipals.  Intuitively, it requires the existence
of a distinguished first-order function $\mathit{sub}_w$ of
type $P \times D_w \to P$ at each world $w$.  Further,
we require that if $sub_w(p,d) = q$, then $A_p \supseteq A_{q}$,
ensuring that super-principals speak for subprincipals.
Since $\mathit{sub}_w$ is a function, it must obey the requirement
of monotonicity w.r.t.\ constructive accessibility relation $\leq$,
just as all other functions $f_{j,w}$ must in a constructive model.
      
\paragraph*{NAL models for Coq.} 
Finally, a \emph{NAL model for Coq} is a NAL model extended with
a pair of sets $\Delta$ and $\Pi$.  This is a technical extension
that unfortunately seems to be necessary in order to express something
that is, in actuality, fairly simple set theory.  We'd like to require
that set $P$ of principals be a subset of every domain
$D_w$ in a NAL model, such that there is one unchanging set
of principals throughout the model.  Expressing that idea
in Coq's type theory turns out to be quite difficult, 
so we instead stipulate the existence of two
sets of coercion functions, $\Delta$ and $\Pi$, between 
principals and individuals.
$\Delta$ is a set $\setdef{\delta_{w} : P \to D_w}{w \in W}$ of 
coercion functions that map principals to individuals.  Since
every principal should be represented by a unique domain element,
we require each $\delta_w$ to be injective.
$\Pi$ is a set $\setdef{\pi_{w} : D_w \to P}{w \in W}$ of 
coercion functions that map individuals to principals.
If individual $d$ does not represent a principal,
then $\delta_w(d)$ is $\bot$.

Given these coercion functions, it is possible to define equality
$=_P$ of principals in terms of equality of individuals:
$p =_P q$ iff for all $w$, $\
\delta_w(p) =_w \delta_w(q)$.

\paragraph*{NAL semantics.} We give a semantics of {\newnal}  
in figure~\ref{fig:semantics}.
The validity judgment is written $$M, w, v \models \phi$$ where
$M$ is a NAL model for Coq and $w$ is a world in that model.
Function $v$ is a \emph{valuation} mapping first-order
variables to individuals; it is used to interpret
first-order quantification.
The semantics also relies on an \emph{interpretation} function $\mu$, 
defined in figure~\ref{fig:meaning}, that maps syntactic
terms $\tau$ to individuals. 

The first-order constructive fragment of the semantics is routine.
The semantics of $\NALSAYS$ follows from the semantics of a $\Box$ modality
in constructive modal logic~\cite{Simpson94,Wijesekera90}. 
The semantics of delegation $\NALSPEAKSFOR$ follows from a standard 
definition in authorization logics~\cite{AbadiBLP93}.
The semantics of restricted delegation is similar to
one presented by Howell~\cite{Howell00}, and
it is a generalization of the semantics of unrestricted
delegation.  (To see this, take $w'''$ in the semantics
of restricted delegation to be the $w''$ from the semantics of
unrestricted delegation.  Then $w'''$ equals $w''$, hence
is in the same equivalence class.)  Restricted delegation
uses an equivalence relation $\equiv_{x:\phi}^w$ on worlds.  
Intuitively, this relation
is used to partition worlds into equivalence classes that agree
on the validity of formula $\phi$ in all valuations, 
assuming the existence of individuals $D_w$.  
Formally, define $w' \equiv_{x:\phi}^w w''$
to hold iff
\begin{align*}
\forall d \in D_w : \quad&(\forall v : M, w', v\subst{d}{x} \models \phi) \\
&\iff (\forall v : M, w'', v\subst{d}{x} \models \phi).
\end{align*}

The interpretation function is also routine, except for the 
interpretation of group principals.  That interpretation
is similar to the algebra of principals defined in the
ABLP logic~\cite{AbadiBLP93}.

\begin{figure}[p]
\begin{equation*}
\renewcommand{\arraystretch}{1.1}
\begin{array}{lcl}
M,w,v \models \nalTrue & & \text{always}  \\
M,w,v \models \nalFalse & & \text{never}  \\
M,w,v \models r_i(\vec{\tau}) & \text{iff} & \mu(M,w,v)(\vec{\tau}) 
  \in r_{i,w} \\
M,w,v \models \tau = \tau' & \text{iff} & \mu(M,w,v)(\tau) =_w
  \mu(M,w,v)(\tau') \\
M,w,v \models \phi_1 \nalAnd \phi_2 & \text{iff} & 
  M,w,v \models \phi_1 \text{~and~}
  M,w,v \models \phi_2 \\
M,w,v \models \phi_1 \nalOr \phi_2 & \text{iff} & 
  M,w,v \models \phi_1 \text{~or~}
  M,w,v \models \phi_2 \\
M,w,v \models \phi_1 \nalImplies \phi_2 & \text{iff} & 
  \text{for all~} w' \geq w: M,w',v \models \phi_1 \\
  & & \text{~implies~}  M,w',v \models \phi_2 \\
M,w,v \models \nalNot \phi & \text{iff} & 
  \text{for all~} w' \geq w: M,w',v \not\models \phi \\
M,w,v \models \nalForall{x}{\phi} & \text{iff} & 
  \text{for all~} w' \geq w, d \in D_{w'}: 
  M,w',v\subst{d}{x} \models \phi \\
M,w,v \models \nalExists{x}{\phi} & \text{iff} & 
  \text{there exists~} d \in D_{w}: 
  M,w,v\subst{d}{x} \models \phi \\
M,w,v \models \nalSays{\tau}{\phi} & \text{iff} & 
  \text{for all~} w', w'' : w \leq w' 
    \text{~and~} (w',w'') \in A_{\mu(M,w,v)(\tau)} \\
   & & \text{~implies~}  M,w'',v \models \phi \\
M,w,v \models \nalSpeaksfor{\tau_1}{\tau_2} & \text{iff} & 
  \text{for all~} w', w'' : (w',w'') \in A_{\mu(M,w,v)(\tau_2)} \\
  & & \text{~implies~} (w',w'') \in A_{\mu(M,w,v)(\tau_1)} \\
M,w,v \models \nalSpeaksforRest{\tau_1}{\tau_2}{x}{\phi} & \text{iff} & 
  \text{for all~} w', w'' : (w',w'') \in A_{\mu(M,w,v)(\tau_2)} \\
   & & \text{~there exists~} w''' : w'' \equiv_{x:\phi}^{w'} w''' \\
   & & \text{~~and~} (w', w''') \in A_{\mu(M,w,v)(\tau_1)} \\
\end{array}
\end{equation*}
\caption{Validity judgment\label{fig:semantics}}
\end{figure}

\begin{figure}[p]
\begin{equation*}
\renewcommand{\arraystretch}{1.1}
\begin{array}{lcl}
\mu(M,w,v)(x) & = & v(x) \\
\mu(M,w,v)(f_j(\vec{\tau})) & = & f_{j,w}(\mu(M,w,v)(\vec{\tau})) \\
\mu(M,w,v)(\nalSubprin{\tau_1}{\tau_2}) & = & 
  \mathit{sub}_w(\mu(M,w,v)(\tau_1), \mu(M,w,v)(\tau_2))\\
\mu(M,w,v)(\nalGroup{x}{\phi}) & = & 
  \gquant{\bigvee}{p}{M,w,v\subst{p}{x} \models \phi}{p} \\
\end{array}
\end{equation*}
\caption{Interpretation function\label{fig:meaning}}
\end{figure}

\section{Soundness}

The soundness theorem for {\newnal} states 
that if $\phi$ is provable from
assumptions $\Gamma$, and that if a model validates all
the formulas in $\Gamma$, then that model must also validate
$\phi$.  Therefore, any provable formula is semantically valid.

\begin{theorem}[Soundness] 
If $\Gamma \proves \phi$ and for all $\psi \in \Gamma$, it holds
that $M,w,v \models \psi$, then $M,w,v \models \phi$.
\end{theorem}

A Coq mechanization of the proof of Soundness is in progress.
Currently, it contains about 3,000 lines of code and implements
all of the proof except for the cases of delegation and restricted
delegation.

The current proof also requires adding an additional assumption as an
axiom: for all $w$ and $w'$, if $w \leq w'$, or 
if there a exists $p$ such that $(w,w') \in A_{p}$, then
it must hold that $\mu(M, w, v)(\tau) = \mu(M, w', v)(\tau)$.
This axiom is actually provable as a theorem for all terms $\tau$ except
for group principals.  Discharging this assumption for
group principals remains an open problem.

%

\section*{Acknowledgments}

Fred B.\ Schneider consulted on the design of the 
proof system and the Kripke semantics.  We thank him,
Mart\'{i}n Abadi, Deepak Garg, and Colin Stirling for
discussions related to this work.  
This work was supported in part by AFOSR grants
F9550-06-0019, FA9550-11-1-0137, and FA9550-12-1-0334,
National Science Foundation grants 0430161, 0964409, 
and CCF-0424422 (TRUST), ONR
grants N00014-01-1-0968 and N00014-09-1-0652, and a 
grant from Microsoft.

\bibliographystyle{plain}
\bibliography{../bib/nal}

\end{document}